\documentclass[twocolumn,pra,aps,showpacs,superscriptaddress]{revtex4}

\usepackage{bbold}
\usepackage{mathptmx}
\usepackage{subfig}
\usepackage{psfrag,graphicx}
\usepackage{dcolumn}
\usepackage{amsmath,amssymb}
\usepackage{bm}
\usepackage{color}
\usepackage{latexsym}
\usepackage{epstopdf}
\usepackage{color}
\usepackage[english]{babel}
\usepackage{latexsym}
\usepackage{psfrag,graphicx}
\usepackage{amsmath}
\usepackage{amssymb}
\usepackage{amsfonts}
\usepackage{bm}
\usepackage{natbib}
\usepackage{epstopdf}
\usepackage{appendix}
\usepackage[utf8]{inputenc}
\usepackage[english]{babel}
\usepackage{aeguill}
\usepackage{ulem}
\usepackage[justification=justified]{caption}

\definecolor{mygrey}{gray}{0.35}
\definecolor{myblue}{rgb}{0.2,0.2,0.8}
\definecolor{myzard}{cmyk}{0,0,0.05,0}
\definecolor{mywhite}{rgb}{1,1,1}
\definecolor{mywhite}{rgb}{1,1,1}
\definecolor{myred}{rgb}{1,0.,0.3}

\usepackage[colorlinks=true,citecolor=myblue,linkcolor=myred]{hyperref}

\def\ba{\begin{align}}
\def\enda{\end{align}}
\def\bi{\begin{itemize}}
\def\ei{\end{itemize}}

\def\be{\begin{equation}}
\def\ee{\end{equation}}
\def\bea{\begin{eqnarray}}
\def\eea{\end{eqnarray}}
\def\bse{\begin{subequations}}
\def\ese{\end{subequations}}

\newcommand{\ket}[1]{|{#1}\rangle}                       % ket
\newcommand{\average}[1]{\langle {#1} \rangle}           % media < >

\newcommand{\Ignore}[1]{ }

\def\i{\text{i}}

\DeclareMathOperator{\sech}{sech}

\begin{document}

\preprint{APS/123-QED}

\title{Classes of exactly solvable Generalized semi-classical Rabi Systems}

\author{R. Grimaudo}
\address{Dipartimento di Fisica e Chimica dell'Universit\`a di Palermo, Via Archirafi, 36, I-90123 Palermo, Italy.}
\address{I.N.F.N., Sezione di Catania, Catania, Italy.}

\author{A. S. M. de Castro}
\affiliation{Universidade Estadual de Ponta Grossa, Departamento de F\'{\i}sica, CEP 84030-900, Ponta Grossa, PR, Brazil.}

\author{H. Nakazato}
\affiliation{Department of Physics, Waseda University, Tokyo 169-8555, Japan.}

\author{A. Messina}%
\address{I.N.F.N., Sezione di Catania, Catania, Italy.}
\address{ Dipartimento di Matematica ed Informatica dell'Universit\`a di Palermo, Via Archirafi, 34, I-90123 Palermo, Italy.}

\date{\today}

\begin{abstract}
The exact quantum dynamics of a single spin-1/2 in a generic time-dependent classical magnetic field is investigated and compared with the quantum motion of a spin-1/2 studied by Rabi and Schwinger.
The possibility of regarding the scenario studied in this paper as a generalization of that considered by Rabi and Schwinger is discussed and a notion of time-dependent resonance condition is introduced and carefully legitimated and analysed.
Several examples help to disclose analogies and departures of the quantum motion induced in a generalized Rabi system with respect to that exhibited by the spin-1/2 in a magnetic field precessing around the $z$-axis.
We find that, under generalized resonance condition, the time evolution of the transition probability $P_+^-(t)$ between the two eigenstates of ${\hat{S}}^z$ may be dominated by a regime of distorted oscillations, or may even exhibit a monotonic behaviour.
At the same time we succeed in predicting no oscillations in the behaviour of $P_+^-(t)$ under general conditions.
New scenarios of experimental interest originating a Landau-Zener transition is brought to light.
Finally, the usefulness of our results is emphasized by showing their applicability in a classical guided wave optics scenario.
%\begin{description}
%\end{description}
\end{abstract}

\pacs{Valid PACS appear here}

\maketitle

%\tableofcontents

\section{Introduction}

Spin language is transversal to every physical scenario wherein, by definition, the states of the specific system under scrutiny live in a finite-dimensional Hilbert space.
When the generally time-dependent Hamiltonian of the system may in particular be expressed as linear combination of the three generators of su(2), the corresponding quantum dynamical problem coincides with that of a spin $j$ in a time-dependent magnetic field.
It is well known that the time evolution of the spin $j$ in this case is fully recoverable from that of a spin $j=1/2$ subjected to the same magnetic field \cite{Hioe}.
On the other hand, however, a closed general form of the 2x2 unitary evolution operator is still unavailable.
As a consequence, to bring to light new dynamical scenarios of a spin-1/2 represents a target of basic and applicative importance in many contexts such as quantum optics \cite{Haroche}, quantum control \cite{Daems, Greilich}, quantum information, and quantum computing \cite{NC,Braak,Oliveira}.

Rabi \cite{Rabi 1937, Rabi 1954} and Schwinger \cite{Schwinger 1937} exactly solved the quantum dynamics of a spin-1/2 in the now called Rabi scenario, that is, subjected to a static magnetic field $B_z$ along the $z$-axis and an r.f. magnetic field rotating in the $x$-$y$ plane with frequency $\omega_{xy}$, namely
\begin{equation} \label{Magn Field Rabi}
\textbf{B}_R(t)= B_\perp[\cos(\omega_{xy} t) \textbf{c}_1-\sin(\omega_{xy} t) \textbf{c}_2]+B_0 \textbf{c}_3,
\end{equation}
$\textbf{c}_1,\textbf{c}_2$ and $\textbf{c}_3$ being fixed unit vectors in the laboratory frame.
Their seminal papers show that the probability of transition between the two Zeeman states generated by $\mathbf{B}_z=B_0\mathbf{c}_3$ is dominated by periodic oscillations reaching maximum amplitude under the so-called resonance condition $\Delta=\omega_{xy}-\omega_L=0$.
Here $\omega_L$ is the spin Larmor frequency.
The exact treatment of this basic problem provides the robust platform for the NMR technology implementation \cite{Vandersypen}.
The recently published special issue on semiclassical and quantum Rabi models \cite{Issue} witnesses the evergreen attractiveness of this problem.

Searching new exactly solvable time-dependent scenarios of a single spin-1/2 could be very interesting and worth both from basic and applicative points of view, especially in the quantum control context.
To pursue this target, over the last years, new methods have been developed to face the problem with an original strategy \cite{Bagrov, KunaNaudts, Das Sarma, Mess-Nak,MGMN}.
We stress that progresses along this direction might be of relevance even for treating time-dependent Hamiltonians describing interacting qubits or qudits \cite{GMN, GMIV, GBNM}.
In turn exact treatments of such scenarios might stimulate the interpretation of experimental results in fields from condensed matter physics \cite{Calvo, Borozdina} to quantum information and quantum computing \cite{Petta, Anderlini, Foletti, Das Sarma Nat}.

In this paper we construct the exact time evolution of a spin-1/2 subjected to time-dependent magnetic fields never considered before.
Our investigation introduces, in a very natural way, three different classes of Generalized Rabi Systems (GRSs) wherein Rabi oscillations of maximum amplitude still survive.
We succeed indeed in identifying generalized resonance conditions which are, as in the Rabi scenario, at the origin of the complete population transfer between the two Zeeman levels of the spin.
We bring to light that, even at resonance, these oscillations might loose its periodic character, significantly differing, thus, from the sinusoidal behaviour occurring in the Rabi scenario.
In this paper we have also considered time-dependent magnetic fields giving rise to exactly solvable models not satisfying the resonance condition.
In this way, we are able to write down a special link among all the time-dependent parameters appearing in the Hamiltonian of the system even if the resonance condition is not met, for which, however, the dynamical problem is exactly solvable.
The departure of the time evolution of the transition probability out of generalized resonance from the corresponding behaviour in the Rabi scenario is illustrated with the help of two exemplary cases.

The paper is organized as follows: in Sec. \ref{MN Res} two sets of parametrized solutions of the dynamical problem of a single spin-1/2 in a time-dependent magnetic field are given.
The generalized time-dependent resonance condition and the generalized out of resonance case are introduced and physically legitimated In Sec. \ref{R and OOR Cases}.
In the subsequent section (Sec. \ref{Examples}) several examples illustrate the occurrence of effects on the Rabi transition probability due to the magnetic field time-dependence.
Finally, in the Conclusion section, a summary of the results with possible applications and future outlooks are briefly discussed.

\section{Resolution of 2x2 su(2) quantum dynamical problems} \label{MN Res}

The problem of a single spin-1/2 subjected to a generic time-dependent magnetic field $\textbf{B}(t) \equiv [B_x(t),B_y(t),B_z(t)]$ is investigated by assuming the su(2) Hamiltonian model
\begin{equation}\label{SU(2) Hamiltonian}
H(t)=
\begin{pmatrix}
\Omega(t) & \omega(t) \\
\omega^*(t) & -\Omega(t)
\end{pmatrix},
\end{equation}
with
\begin{subequations} \label{Def Magn Field}
\begin{align}
&\Omega(t)={\hbar \mu_0 g \over 2} B_z(t), \\
&\omega(t)={\hbar \mu_0 g \over 2} [B_x(t)-iB_y(t)] \equiv \omega_x-i \omega_y \equiv |\omega(t)|e^{i\phi_\omega(t)}.
\end{align}
\end{subequations}
Here $\mu_0 g$ is the magnetic moment associated to the spin-1/2, $g$ and $\mu_0$ being the appropriate Land\'e factor and the Bohr magneton, respectively.
The entries $a(t)\equiv|a(t)|\exp \{i \phi_a(t)\}$ and $b(t)\equiv|b(t)|\exp \{i\phi_b(t)\}$ of the unitary time evolution operator
\begin{equation} \label{Time Ev Op SU2}
U (t)=
\begin{pmatrix}
a(t) & b(t) \\
-b^*(t) & a^*(t)
\end{pmatrix}, \quad
|a(t)|^2+|b(t)|^2=1,
\end{equation}
generated by $H(t)$, must satisfy the Cauchy-Liouville problem $i\hbar\dot{U}(t)=H(t)U(t)$, $U(0)=\mathbb{1}$, which originates the following system of linear differential equations
\begin{equation}\label{LC problem}
  \left\{
  \begin{aligned}
  &\dot{a}(t)=\frac{\Omega}{i\hbar}a(t)-\frac{\omega}{i\hbar}b^{*}(t),\\
  &\dot{b}(t)=\frac{\omega}{i\hbar}a^{*}(t)+\frac{\Omega}{i\hbar}b(t),\\
  &a(0)=1,\quad b(0)=0.
  \end{aligned}
  \right.
\end{equation}

It is possible to demonstrate \cite{Mess-Nak} that if $\Theta(t)$ is a complex-valued $C1$ function of $t$ satisfying the nonlinear integral-differential Cauchy problem
\begin{subequations}\label{Cauchy problem Theta}
\begin{align}
&{1\over2}\dot{\Theta}(t) + {|\omega(t)| \over \hbar}\sin\Theta(t)
\cot\Bigl[{2\over\hbar} \int_0^t|\omega(t')|  \cos\Theta(t') dt'\Bigr]= \nonumber \\
&={\Omega(t) \over \hbar}+{\dot{\phi}_\omega(t) \over 2}, \label{Rel exactly solvable scenario gen} \\
&\Theta(0)=0,
\end{align}
\end{subequations}
then the solutions of the Cauchy problem \eqref{LC problem} can be represented as follows
\begin{subequations}\label{a and b}
\begin{align}
a(t) =& \cos\biggl[{ 1 \over \hbar}\int_0^t |\omega(t')|\cos \bigl[ \Theta(t') \bigr] dt' \biggr] \times \nonumber \\
&\exp\left\{i\left(\dfrac{\phi_\omega(t)-\phi_\omega(0)}{2} - {\Theta(t) \over 2} - \mathcal{R}(t)\right) \right\}, \label{Gen a} \\
b(t) =& \sin\biggl[{1\over\hbar}\int_0^t|\omega(t')|\cos \bigl[ \Theta(t') \bigr] dt' \biggr] \times \nonumber \\
&\exp\left\{ i\left(\dfrac{\phi_\omega(t)+\phi_\omega(0)}{2} - {\Theta(t) \over 2} + \mathcal{R}(t) - \dfrac{\pi}{2}\right) \right\}, \label{Gen b}
\end{align}
\end{subequations}
with
\begin{equation}\label{Integral R}
\mathcal{R}(t)=\int_0^t \dfrac{|\omega(t')|\sin[\Theta(t')]}{\sin\left[ 2 \int_0^{t'} |\omega(t'')|\cos[\Theta(t'')]dt'' \right]}dt'.
\end{equation}
\textit{Vice versa}, if $a(t)$ and $b(t)$ are solutions of the Cauchy problem \eqref{LC problem}, then the representations given in Eqs. \eqref{a and b} are still valid and $\Theta(t)$ satisfies Eqs. \eqref{Cauchy problem Theta}.

Generally speaking, solving Eq. \eqref{Cauchy problem Theta} is a difficult task. %when no analytical connection between the time-dependence of $B_z(t)$ and that of $B_x(t)-iB_y(t)$ has been postulated.
This equation however may be exploited in a different way, giving rise to a strategy \cite{Mess-Nak} aimed at singling out exactly solvable dynamical problems represented by Eq. \eqref{LC problem}.
Fixing, indeed, at will the function $\Theta(t)$ in Eqs. \eqref{Cauchy problem Theta}, that is, $\Theta(t)$ regarded now as a parameter (function) rather than an unknown, determines a link between $\Omega (t)$ and $\omega (t)$ under which the corresponding dynamical problem may be exactly solved in view of Eqs. \eqref{a and b}.
We emphasize that if we knew the solution of the Cauchy problem given in Eq. \eqref{Cauchy problem Theta}, whatever $\Omega(t)$, $|\omega(t)|$ and $\dot{\phi}_\omega(t)$ are, then we would be in condition to solve in general the corresponding Cauchy dynamical problem expressed by Eqs. \eqref{LC problem}.

Another useful way of parametrizing the expressions of $a(t)$ and $b(t)$ is
\begin{subequations}\label{a and b particular}
\begin{align}
a(t)=&\sqrt{\hbar^2+c^2\cos^2[\Phi(t)]\over\hbar^2+c^2} \times \nonumber \\
&\exp\left\{ i\left( {\phi_\omega(t)\over 2}-\tan^{-1}\left[{\hbar\over\sqrt{\hbar^2+c^2}}\tan[\Phi(t)]\right] \right) \right\},\\
b(t)=&{c\over\sqrt{\hbar^2+c^2}}\sin[\Phi(t)]
\exp\left\{ i\left( {\phi_\omega(t) \over 2}-{\pi \over 2} \right) \right\}, \label{b special cond}
\end{align}
\end{subequations}
with
\begin{equation}
\Phi(t)={\sqrt{\hbar^2+c^2}\over \hbar c} \int_0^t|\omega(t')|dt',
\end{equation}
$c$ being an arbitrary real number and having put, without loss of generality, $\phi_\omega(0)=0$.
In this case, it is possible to check that they solve the system \eqref{LC problem} if the following condition holds
\begin{equation}\label{Relation exactly solvable scenario}
{|\omega(t)| \over c}={\Omega(t) \over \hbar}+{\dot{\phi}_\omega(t) \over 2}.
\end{equation}
It is stressed that this last equation does only express the condition under which, whatever $c$ is, the representations \eqref{a and b particular} satisfy the Cauchy problem \eqref{LC problem}.
This means that the real number $c$ plays in this case the role of parameter.
When Eq. \eqref{Relation exactly solvable scenario} cannot be satisfied for any $c$, of course the solution of the dynamical problem exists but cannot be represented using Eqs. \eqref{a and b particular}.
In this case there certainly exists a function $\Theta(t)$ enabling the representation of the solutions by using Eqs. \eqref{a and b}.
Finally, it is interesting to underline that Eq. \eqref{Rel exactly solvable scenario gen} turns into the simpler condition \eqref{Relation exactly solvable scenario} on $\textbf{B}(t)$ under an appropriate choice of the parameter function $\Theta(t)$ \cite{Mess-Nak}.

\section{Generalized Resonance Condition and out of Resonance Cases}\label{R and OOR Cases}

The experimental set-up considered by Rabi, as described in the introduction, leads to the Hamiltonian model \eqref{SU(2) Hamiltonian} where
\begin{subequations} \label{Def Magn Field}
\begin{align}
&\Omega(t)={\hbar \mu_0 g \over 2} B_0 \equiv \Omega_0, \\
&|\omega(t)|={\hbar \mu_0 g \over 2} \sqrt{B_x^2(t)+B_y^2(t)}={\hbar \mu_0 g \over 2}B_\perp \equiv |\omega_0|,\\
&\phi_\omega(t)=\nu_0 t \equiv \dot{\phi}_0 t.
\end{align}
\end{subequations}
Then it is characterized by the three time-independent parameters: $\Omega_0$, $|\omega_0|$ and $\dot{\phi}_0$.
In this paper we generalize this Rabi scenario by making some out of or all these parameters time-dependent: $\Omega \rightarrow \Omega(t)$, $|\omega| \rightarrow |\omega(t)|$ and $\dot{\phi}_\omega \rightarrow \dot{\phi}_\omega(t)$.

Firstly, we rewrite the general Hamiltonian \eqref{SU(2) Hamiltonian} as follows
\begin{equation}
H=\Omega(t) \hat{\sigma}^z + \omega_x(t) \hat{\sigma}^x + \omega_y(t) \hat{\sigma}^y,
\end{equation}
with $\omega_{x/y}(t)=\hbar\mu_0gB_{x/y}(t)/2$ and $\hat\sigma^{x/y/z}$ being Pauli matrices.
Generalizing the approach in Ref. \cite{Rabi 1954}, we pass from the laboratory frame to the time-dependent one tuned with $\phi_\omega(t)$, where the time-dependent Schr\"odinger equation for the transformed state,
\begin{equation}
\ket{\psi(t)}=\exp\{i \phi_\omega(t) \hat{\sigma}^z/2\} \ket{\tilde{\psi}(t)},
\end{equation}
is governed by the following effective time-dependent transformed Hamiltonian
\begin{equation}\label{H eff}
H_{GR}(t)=\left( \Omega(t)+{\hbar \over 2}\dot{\phi}_\omega(t) \right) \hat{\sigma}^z + |\omega(t)| \hat{\sigma}^x.
\end{equation}

It is worth noticing its strict similarity with the analogous one got in Ref. \cite{Rabi 1954} where the unitary transformation is indeed a uniform rotation around the $z$-axis.
In fact, it is enough to make $\Omega(t)$, $|\omega(t)|$ and $\dot{\phi}_\omega(t)$ time-independent in $H_{GR}$ ($GR$ stands for Generalized Rabi) to immediately recover the transformed Hamiltonian got by Rabi \cite{Rabi 1954}.
On the basis of this observation it then appears natural to refer to the following condition
\begin{equation}\label{Gen Res Rabi Cond}
{\Omega(t)}+{\hbar \over 2}\dot{\phi}_\omega(t)=0,
\end{equation}
as a generalized resonance condition, in accordance with the corresponding static resonance condition ${\Omega_0}+{\hbar\dot{\phi}_0 / 2}=0$ brought to light by Rabi in Ref. \cite{Rabi 1937}.
We underline that the generalized resonance condition does not lead to a time-independent transformed dynamical problem (as it happens in the Rabi scenario), but, whatever $H$ is, it easily enables the explicit construction of the time evolution operator describing the quantum motion of the spin in the laboratory frame.
In view of Eq. \eqref{Relation exactly solvable scenario}, the entries of such an operator are indeed exactly given by Eqs. \eqref{a and b particular} in the limit $c \rightarrow \infty$, namely
\begin{subequations}\label{a and b c inf}
\begin{align}
a(t)=&\cos\biggl[ \int_0^t {|\omega| \over \hbar} dt' \biggr] \text{exp}\Bigl\{i{\phi_\omega(t)\over 2} \Bigr\}, \\
b(t)=&\sin\biggl[ \int_0^t {|\omega| \over \hbar} dt' \biggr]\text{exp}\left\{i{\phi_\omega(t)\over 2}-i{\pi \over 2}\right\}.
\end{align}
\end{subequations}

By definition, we say to be in generalized out of resonance when the left hand side of Eq. \eqref{Gen Res Rabi Cond} is non-vanishing, namely
\begin{equation}\label{OoRC}
{\Omega(t) \over \hbar}+{\dot{\phi}_\omega(t) \over 2}=\Delta(t) \neq 0,
\end{equation}
where $\Delta(t)$ is an arbitrary energy-dimensioned well-behaved function of time.
Generally speaking, to find the exact quantum dynamics of the spin in the generic out of resonance case, is a very complicated mathematical problem.
Let us observe that, on the basis of the structure of $H_{GR}$ in Eq. \eqref{H eff}, when $\Delta(t)$ is proportional to $|\omega(t)|$, the dynamical problem may be exactly solved.
Indeed, this condition coincides with that expressed by Eq. \eqref{Relation exactly solvable scenario} which in turn enables one to write down exact solutions of the Cauchy problem \eqref{LC problem} in the form given by Eqs. \eqref{a and b particular}.
In this paper we report the exact solutions of special non trivial out of resonance dynamical problems.
Our aim is to illustrate the occurrence of analogies and differences in the time behaviour on the Rabi transition probability
\begin{equation}
P_+^-(t)=|\average{-|U(t)|+}|^2=|b(t)|^2,
\end{equation}
($\hat{\sigma}^z\ket{\pm}=\pm\ket{\pm}$), when the time evolution of the magnetic field acting upon the spin cannot be described as a perfect precession around the $z$-axis.

\section{Examples of Generalized Rabi models} \label{Examples}

This section is aimed at showing how the Rabi transition probability $P_+^-(t)=|\average{-|U(t)|+}|^2$ is only slightly as well as strongly affected by different choices of the time-dependent magnetic fields under general conditions.
The following examples are reported to illustrate such behaviour.

\subsection{Examples of GRSs Dynamics under Generalized Resonance Condition}\label{Var One Par}

\begin{figure}[htp]
\centering
%\subfloat[][]{\includegraphics[scale=.4]{MagFDampedx.eps}\label{fig:MFX}}
%\qquad
%\subfloat[][]{\includegraphics[scale=.4]{MagFDampedy.eps}\label{fig:MFY}}
%\qquad \\
\subfloat[][]{\includegraphics[scale=.4]{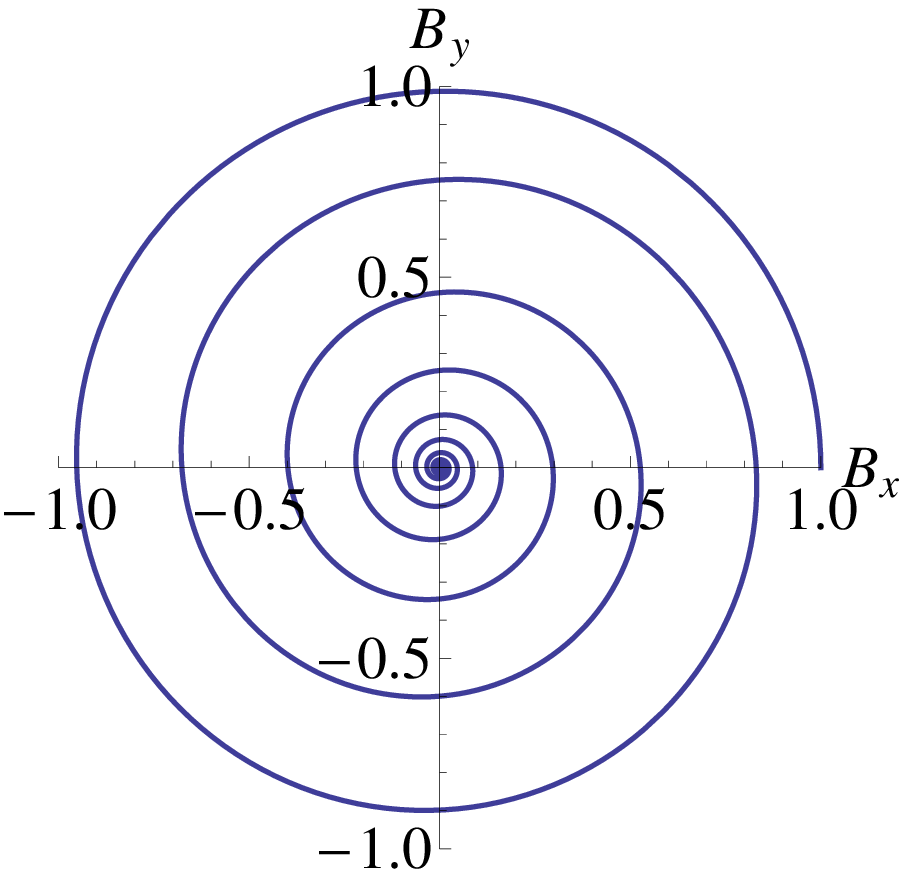}\label{fig:MFMP}}
\qquad
\subfloat[][]{\includegraphics[scale=.4]{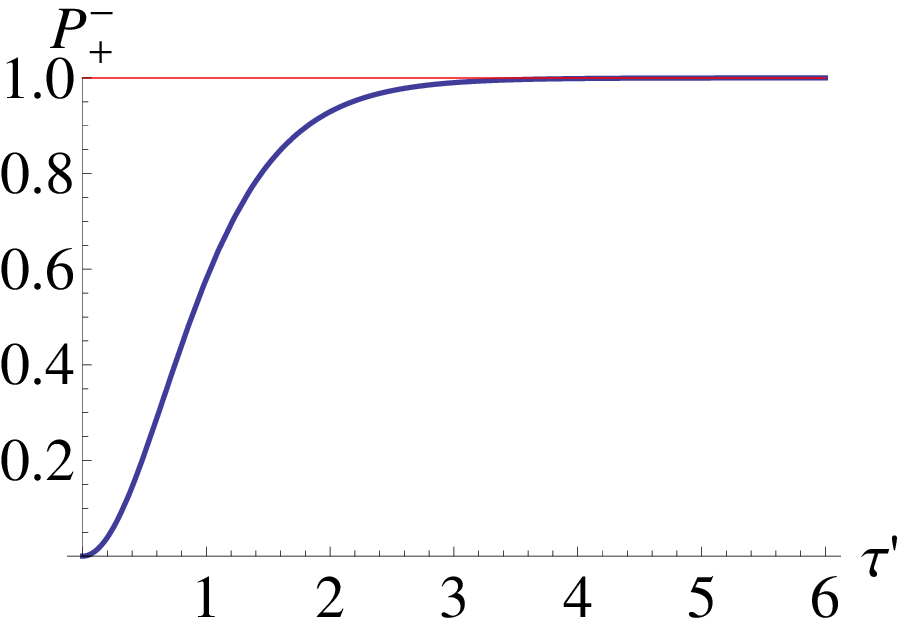}\label{fig:MP}}
\\
\subfloat[][]{\includegraphics[scale=.4]{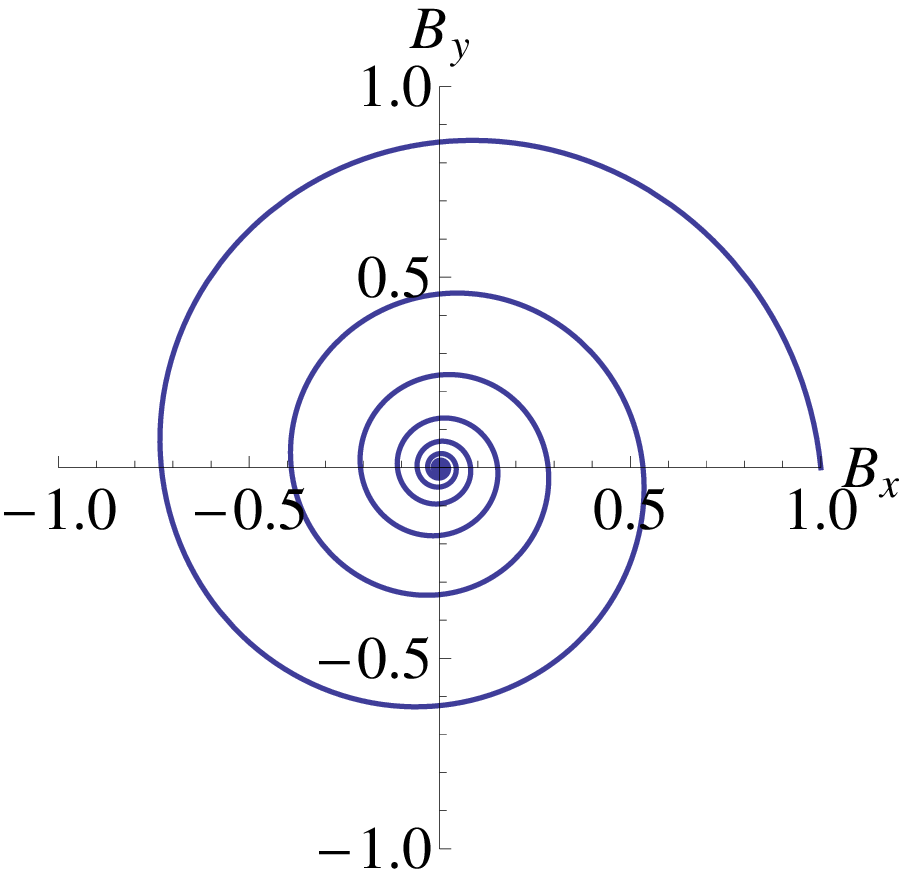}\label{fig:MFDXY}}
\qquad
\subfloat[][]{\includegraphics[scale=.4]{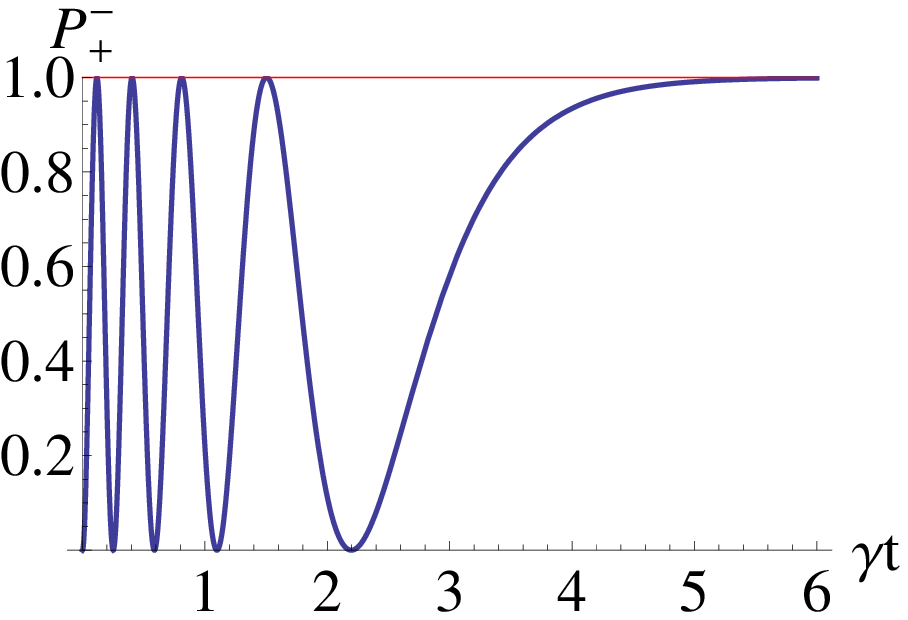}\label{fig:PMFDXY}}
\\
\subfloat[][]{\includegraphics[scale=.4]{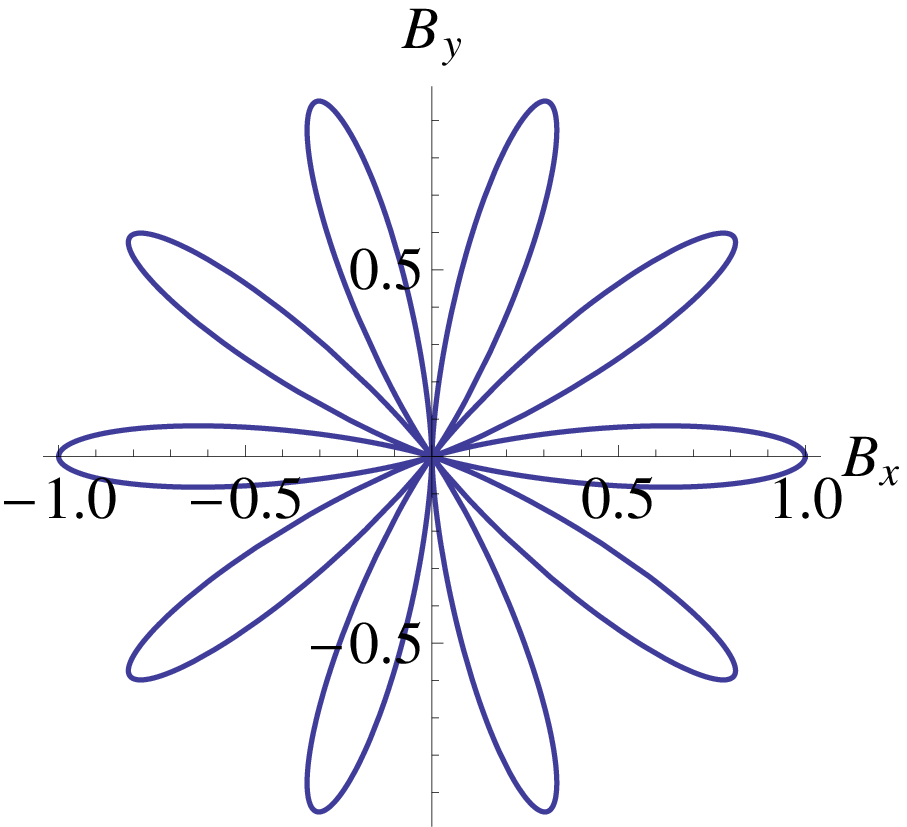}\label{fig:TFS}}
\qquad
\subfloat[][]{\includegraphics[scale=.4]{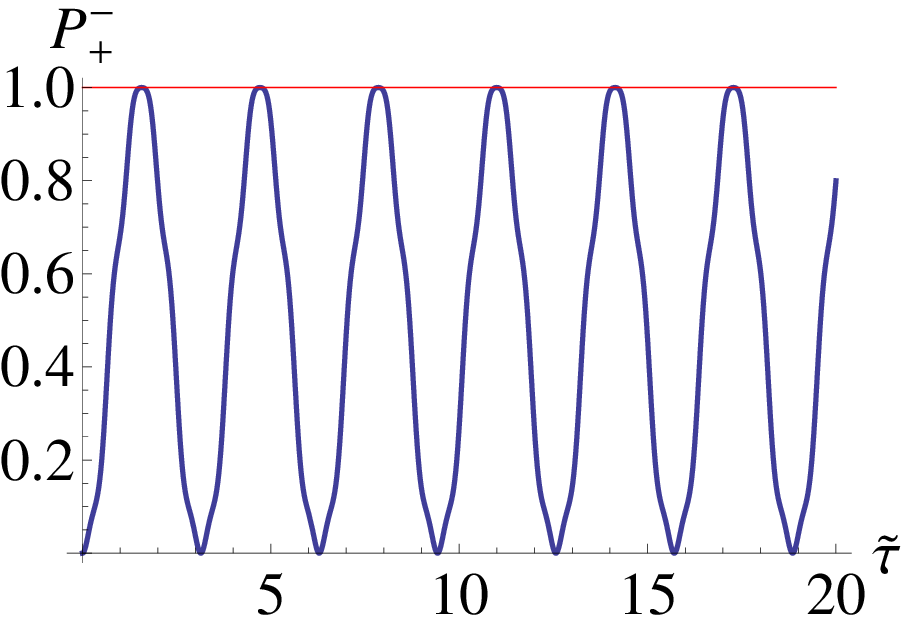}\label{fig:PTFS}}
\captionsetup{justification=justified,format=plain,skip=4pt}%
\caption{(Color online)
a) The normalized magnetic field in Eq. \eqref{Transverse mag field sech decreasing}, parametrically represented in the $x$-$y$ plane and the joined b) transition probability in Eq. \eqref{Trans Prob sech decreasing} against the dimensionless parameter $\tilde{\tau}=|\omega_0|t/\hbar$ with $\dot{\phi}_0/\hbar|\omega_0|=10$;
%Time-dependence of the magnetic field components in Eq. \eqref{Transverse mag field exp decreasing} against $\gamma t$ with ${|\omega_0| / \hbar \gamma}={9 \pi/ 2}$ and $\dot{\phi}_0/\gamma=10$. Figures a) and b) report the $x$ and $y$ component, respectively; figure
c) the normalized magnetic field in Eq. \eqref{Transverse mag field exp decreasing}, %$(B_{x}(t)\mathbf{c}_1+B_{y}(t)\mathbf{c}_2) / B_\perp$,
parametrically represented in the $x$-$y$ plane and the related d) transition probability in Eq. \eqref{Trans Prob alfa} against the dimensionless parameter $\gamma t$ with ${|\omega_0| / \hbar \gamma}={9 \pi/ 2}$ and $\dot{\phi}_0/\gamma=10$;
e) the normalized transverse magnetic field %$[B_{x}(t)\mathbf{c}_1+B_{y}(t)\mathbf{c}_2] / (B_\perp+A')$, with $B_{x}(t)$ and $B_{y}(t)$
in Eq. \eqref{Mag Field Somm}, parametrically represented in the $x$-$y$ plane in terms of $\tilde{\tau}=\dot{\phi}_0 t$ with ${A' / B_\perp}=1$ and $\lambda=10\dot{\phi}_0$ and the related f) transition probability in Eq. \eqref{Trans Prob TFS Parametrized} for $k=1$ and $n=10$ and $C=1$.}
\end{figure}

Let us consider, firstly, the generalized resonance condition in Eq. \eqref{Gen Res Rabi Cond}.
We know that, in this instance, the time evolution operator is characterized by the time behaviour of its two entries given in Eq. \eqref{a and b c inf}, so that the transition probability reads
\begin{equation}\label{P+- Res Cond}
P_+^-(t)=\sin^2\biggl[ \int_0^t {|\omega| \over \hbar} dt' \biggr].
\end{equation}
It is immediately evident that $P_+^-(t)$ could or could not be periodic.
Indeed, e.g., setting $|\omega(t)|=|\omega_0|\sech({|\omega_0|t/\hbar})$, obtainable by an $x$-$y$ magnetic field varying over time as
\begin{equation} \label{Transverse mag field sech decreasing}
\begin{aligned}
\textbf{B}_{tr} =& B_x(t) \textbf{c}_1 + B_y(t) \textbf{c}_2 \\
=& B_\perp \sech({|\omega_0|t/\hbar}) [\cos\left(\dot{\phi}_0 t\right) \textbf{c}_1 - \sin\left(\dot{\phi}_0 t\right)\textbf{c}_2],
\end{aligned}
\end{equation}
we get
\begin{equation}\label{Trans Prob sech decreasing}
P_+^-(t)=\tanh^2(|\omega_0|t/\hbar),
\end{equation}
resulting in a Landau-Zener-like transition \cite{LZ}, that is an asymptotic aperiodic inversion of population.
Figures \ref{fig:MFMP} and \ref{fig:MP} represent the transverse magnetic field in Eq. \eqref{Transverse mag field sech decreasing} and the resulting transition probability in Eq. \eqref{Trans Prob sech decreasing}, respectively, plotted against the dimensionless time $\tilde{\tau}=|\omega_0|t/\hbar$ with $\dot{\phi}_0/\hbar|\omega_0|=10$.

However, of course, it is easy to understand that it is possible to make choices either resulting in a oscillating but not periodic transition probability or exhibiting a periodic behaviour, even if not coincident with that characterizing the Rabi scenario.
If we consider, for example,
\begin{equation}\label{Mod omega exp}
|\omega(t)|=|\omega_0|e^{-\gamma t},
\end{equation}
reproducible by engineering the transverse magnetic field as
\begin{equation} \label{Transverse mag field exp decreasing}
\begin{aligned}
\textbf{B}_{tr} =& B_x(t) \textbf{c}_1 + B_y(t) \textbf{c}_2 \\
=& B_\perp e^{-\gamma t} [\cos\left(\dot{\phi}_0 t\right) \textbf{c}_1 - \sin\left(\dot{\phi}_0 t\right)\textbf{c}_2],
\end{aligned}
\end{equation}
the resulting transition probability yields
\begin{equation}\label{Trans Prob alfa}
P_+^-(t)=\sin^2[\alpha(1-e^{-\gamma t})],
\end{equation}
with $\alpha={|\omega_0| / \hbar \gamma}$.
We point out that, for the sake of simplicity, in Eqs. \eqref{Transverse mag field sech decreasing} and \eqref{Transverse mag field exp decreasing} we have put $\phi_\omega(t)=\dot{\phi}_0 t$, even if, in general, the expression of the probability in Eq. \eqref{Trans Prob alfa} holds whatever $\phi_\omega(t)$ is, provided that Eq. \eqref{Gen Res Rabi Cond} is satisfied.
%In Figs. \ref{fig:MFX} and \ref{fig:MFY} we report the time dependence of the two components of the transverse magnetic field, with respect to $B_\perp$, accomplishing the time dependence of $|\omega(t)|$ given in Eq. \eqref{Mod omega exp}.
Figure \ref{fig:MFDXY} shows the time behaviour of the magnetic field in the $x$-$y$ plane, against the dimensionless parameter $\gamma t$, when $\alpha={9 \pi/ 2}$ and $\dot{\phi}_0/\gamma=10$.

The time behaviour of $P_+^-(t)$ as given in Eq. \eqref{Trans Prob alfa} is reported in Fig. \ref{fig:PMFDXY} for $\alpha=9\pi/2 $.
%\begin{figure}[htbp]
%\centering
%\includegraphics[scale=1]{P+-MFD.eps}
%\caption{(Color online) Time-dependence of the transition probability $P_+^-(t) \equiv \average{-|U(t)|+}$ relative to the scenario characterized by the time-dependent magnetic field in Eq. \eqref{Transverse mag field exp decreasing} with $\alpha=|\omega_0| / \hbar \gamma={9 / 2}\pi$; the upper straight line represents $P_+^-=1$.} \label{fig:P+-MFD}
%\end{figure}
We recognize the existence of a transient wherein $P_+^-(t)$ exhibits aperiodic oscillations of maximum amplitude which, after a finite interval of time, turn into a monotonic increase that asymptotically approaches 1.
We emphasize that the number of complete oscillations, preceding the asymptotic behaviour of $P_+^-(t)$ as well as $P_+^-(\infty)$ itself, are $\alpha$-dependent.
Equation \eqref{Trans Prob alfa}, indeed, predicts
\begin{equation}
P_+^-(\infty)=\sin^2(\alpha),
\end{equation}
which immediately leads to
\begin{equation}
\left\{
\begin{aligned}
&P_+^-(\infty)=0, \quad \alpha=n\pi, \\
&P_+^-(\infty)=1, \quad \alpha={2n+1\over 2}\pi, \\
&P_+^-(\infty)=\sin^2(\alpha), \quad \text{otherwise}.
\end{aligned}
\right.
\end{equation}

As our third example, we consider the following modulation of $|\omega(t)|$
\begin{equation}\label{Choice of mod omega}
|\omega(t)|=|\omega_0|+A\cos(\lambda t),
\end{equation}
realizable by engineering the transverse magnetic field as
\begin{equation}\label{Mag Field Somm}
\begin{aligned}
B_x(t)&=[B_\perp+A'\cos(\lambda t)]\cos(\dot{\phi}_0 t), \\
B_y(t)&=-[B_\perp+A'\cos(\lambda t)]\sin(\dot{\phi}_0 t).
\end{aligned}
\end{equation}
Here $A={\hbar\mu_0gA' / 2}$, $A'>0$ and $\lambda={n\dot{\phi}_0}$ with $n \in \mathbb{N}^*$.
The transverse field is represented in Fig. \ref{fig:TFS} against the adimensional time parameter $\tilde{\tau}=\dot{\phi}_0 t$, once more supposing for simplicity $\phi_\omega(t)=\dot{\phi}_0 t$.
%\begin{figure}[htbp]
%\centering
%\includegraphics[scale=1]{TransFieldSomm.eps}
%\caption{(Color online)
%Time behaviour of the normalized transverse magnetic field $[B_{x}(t)\mathbf{c}_1+B_{y}(t)\mathbf{c}_2] / (B_\perp+A')$, with $B_{x}(t)$ and $B_{y}(t)$ in Eq. \eqref{Mag Field Somm}, parametrically represented in the $x$-$y$ plane in terms of $\tilde{\tau}=\dot{\phi}_0 t$. We have set ${A' / B_\perp}=1$ and $\lambda=10\dot{\phi}_0$.} \label{TFS}
%\end{figure}

In this case, the Rabi's transition probability results
\begin{equation}\label{Trans Prob TFS Parametrized}
P_+^-(t)=\sin^2\Bigl[ C\Bigl(\tilde{\tau} + {k \over n} \sin(n\tilde{\tau}) \Bigr) \Bigr],
\end{equation}
with
\begin{equation}
C={|\omega_0| \over \hbar \dot{\phi}_0}, \quad k={A' \over B_\perp}, \quad \tilde{\tau}=\dot{\phi}_0 t, \quad n={\lambda \over \dot{\phi}_0}.
\end{equation}
The behaviour of $P_+^-(t)$ in Eq. \eqref{Trans Prob TFS Parametrized} is shown in Fig. \ref{fig:PTFS}, having put $k=1$, $n=10$ and $C=1$.
%\begin{figure}[htbp]
%\centering
%\includegraphics[scale=1]{Modb2TFS.eps}
%\caption{(Color online) Plot of the transition probability in Eq. \eqref{Trans Prob TFS Parametrized} for $k=1$ and $n=10$ and $C=1$.} \label{Mb2TFS}
%\end{figure}
Differently from the previous example, we see that, in this case, the characteristic sinusoidal behaviour of the Rabi transition probability turns into a periodic population transfer, still of maximum amplitude, between the two energy levels of the spin.
We emphasize that, in view of Eq. \eqref{P+- Res Cond}, different time evolutions of $P_+^-(t)$ require different choices of $|\omega(t)|$ only, then regardless of $\Omega(t)$ and $\phi_\omega(t)$ which are constrained by the generalized resonance condition \eqref{Gen Res Rabi Cond} only.
We stress, however, that distinct realizations of the resonance condition, keeping the same $|\omega(t)|$, introduce significant changes in the dynamical behaviour of the GRS with respect to the Rabi system.
It is enough to consider, for example, that
\begin{equation}
\average{+|U^\dagger(t)\hat{\sigma}^{x/y}U(t)|+}=\mp 2 \hbar |a(t)||b(t)| \cos[\phi_a(t)+\phi_b(t)],
\end{equation}
depend on both $\phi_\omega(t)$ and $|\omega(t)|$, in view of Eqs. \eqref{a and b c inf}.

\subsection{Examples of GRSs Dynamics in Generalized out of Resonance Cases}\label{Var Two Par}

In this subsection we analyse the generalized out of resonance case, defined in Eq. \eqref{OoRC}.
Since it appears hopeless to have an exact closed treatment of the Cauchy problem in Eq. \eqref{Cauchy problem Theta} with an arbitrary $\Delta(t)$, we confine ourselves to the following specific forms
\begin{equation}
\hbar\Delta(t)=
\left\{
\begin{aligned}
&\beta_0 |\omega(t)|, \\
&\beta(t) |\omega(t)|.
\end{aligned}
\right.
\end{equation}
where $\beta_0$ and $\beta(t)$ are non-negative adimensional functions.
The first form coincides with the condition in Eq. \eqref{Relation exactly solvable scenario} with $\beta_0=\hbar/c$.
In this case, the solutions $a(t)$ and $b(t)$ of the system in Eq. \eqref{LC problem} may be cast as reported in Eqs. \eqref{a and b particular} so that
\begin{equation}\label{P+- OoR c}
P_+^-(t)={1\over1+\beta_0^2}\sin^2\left[ \sqrt{1+\beta_0^2} \int_0^t{|\omega(t')|\over \hbar}dt' \right].
\end{equation}
In the limit $\beta_0\rightarrow 0$ we recover Eq. \eqref{P+- Res Cond} from this equation.
Thus, we may compare $P_+^-(t)$ in the resonant and this non-resonant cases when $|\omega(t)|$ is fixed in the same way.
It is easy to convince oneself that the main effect of a positive value of the parameter $\beta_0$ on $P_+^-(t)$ is nothing but a scale effect determined by the ratio $1/(1+\beta_0^2)$.

We wish now to bring to light and to discuss some exactly solvable scenarios of generalized, out of resonance, Rabi problems wherein $\Delta(t)=\beta(t)|\omega(t)|$.
To this end, it is useful to observe that postulating $\Theta(t)$ as function of $t$ through
\begin{equation}\label{tau}
\tau(t) = \int_0^t{|\omega(t')| \over \hbar}dt',
\end{equation}
we would get, by Eq. \eqref{Cauchy problem Theta}, the desired form of $\Delta(t)$, by construction.
We stress however that the corresponding function $\beta(t)$ would be functionally-dependent on $|\omega(t)|$, that is determined by the knowledge of $|\omega(t)|$.
We emphasize that this aspect, however, does not spoil of interest such a particular procedure.
In the following examples we indeed report two applications of the general strategy \cite{Mess-Nak} exposed after Eq. \eqref{Integral R} in Sec. \eqref{MN Res}, where those $\beta(t)$s that make Eqs.\ (6) solvable are fixed.

\subsubsection{CASE 1}\label{Case 1}
\begin{figure}[htp]
\centering
\subfloat[][]{\includegraphics[scale=.4]{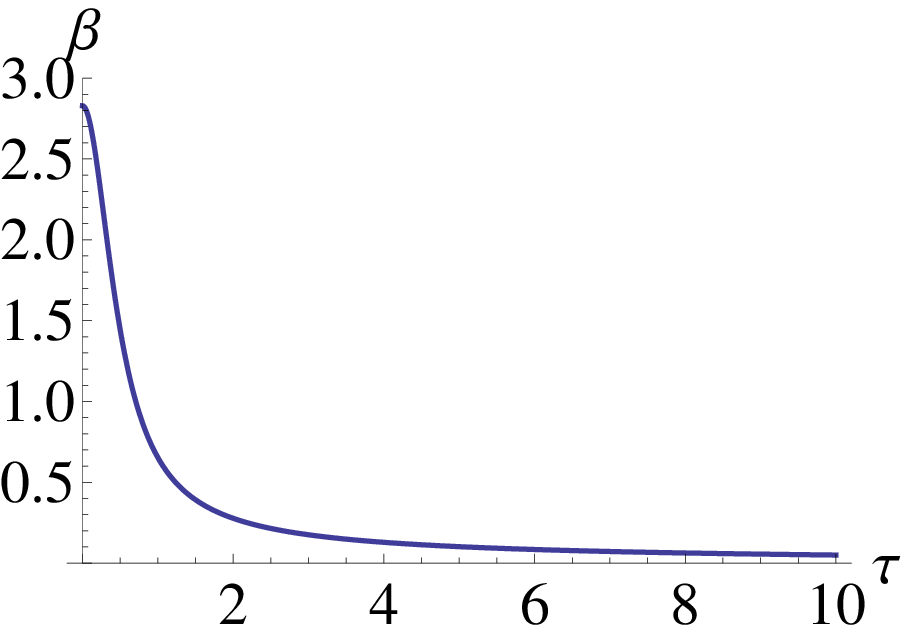}\label{fig:DC1}}
\qquad
\subfloat[][]{\includegraphics[scale=.4]{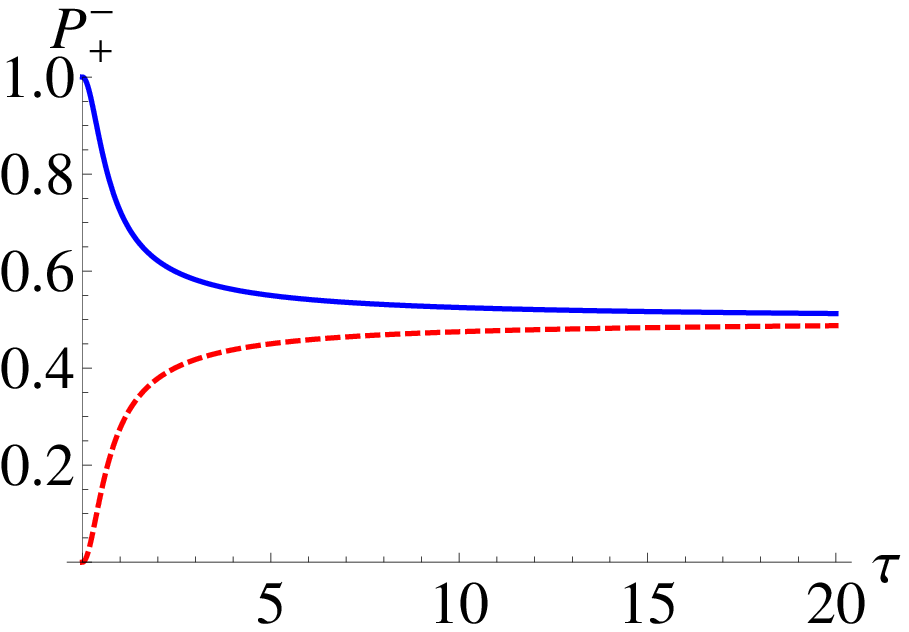}\label{fig:TPB1}}
\\
\subfloat[][]{\includegraphics[scale=.4]{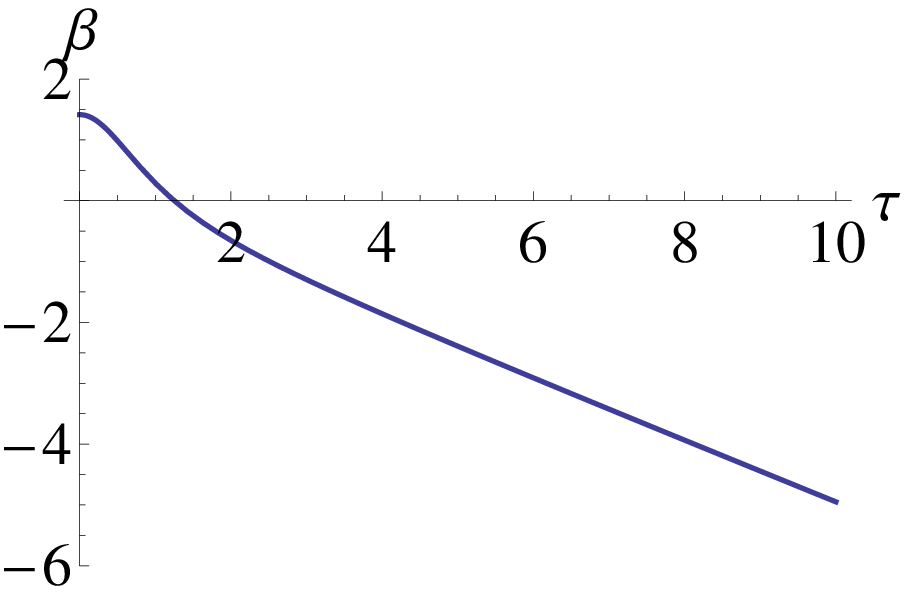}\label{fig:DC2}}
\qquad
\subfloat[][]{\includegraphics[scale=.4]{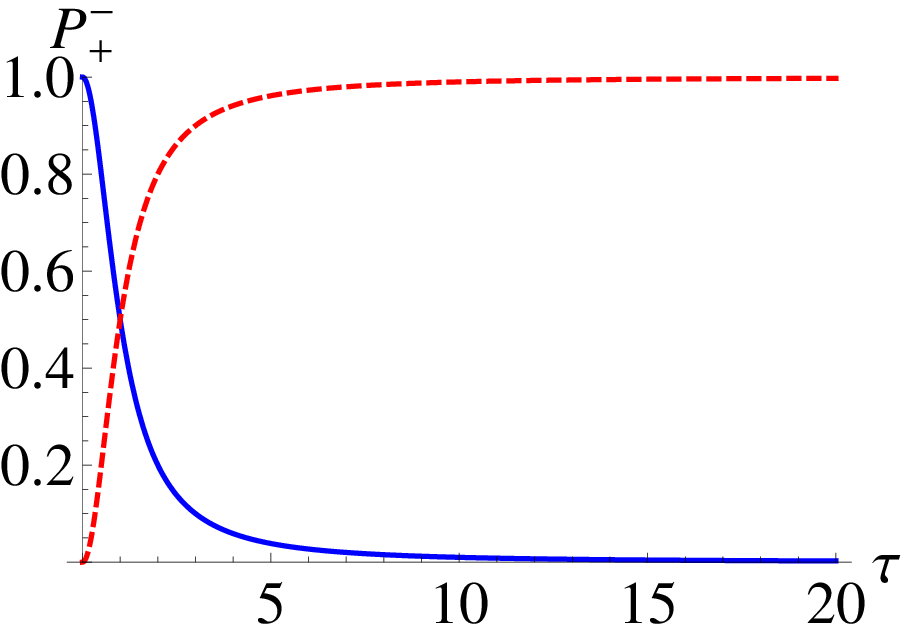}\label{fig:TPB2}}
\captionsetup{justification=justified,format=plain,skip=4pt}%
\caption{(Color online) Time-dependence of the ``detuning'' $\hbar\Delta(\tau)/|\omega_0|=\beta(\tau)$ against the adimensional time $\tau=|\omega_0|t/\hbar$ related to the example a) \ref{Case 1} and c) \ref{Case 2};
transition probabilities $P_+^+(\tau)=|a(\tau)|^2$ (full blue) and $P_+^-(\tau)=|b(\tau)|^2$ (dashed red) for the time-dependent scenario b) \ref{Case 1} and d) \ref{Case 2}.
}
\end{figure}

Assuming the solution of the Cauchy problem \eqref{Cauchy problem Theta} as
\begin{align}\label{Theta 1}
\Theta(t) = 2 \tan^{-1}\left({2\tau \over \sqrt{2+4\tau^2}}\right),
\end{align}
it is straightforward to show that
\begin{equation} \label{Position 3}
\int_0^t{|\omega(t')|\over\hbar}\cos[\Theta(t')] dt' = {1 \over 2} \tan^{-1}(2\tau).
\end{equation}
Equation \eqref{Rel exactly solvable scenario gen} immediately yields
\begin{equation} \label{Omega case 3}
\Delta(t)= {4(1+\tau^2) \over (1+4\tau^2) \sqrt{2+4\tau^2}} {|\omega(t)| \over \hbar} \equiv \tilde{\beta}(\tau)|\omega(t)|=\beta(t)|\omega(t)|.
\end{equation}
From this point on, we are ready to specialize Eqs. \eqref{a and b} getting
\begin{equation}\label{a b case 3}
|a(t)| = \sqrt{{\sqrt{1+4\tau^2} + 1 \over 2 \sqrt{1+4\tau^2}}}, \quad
|b(t)| = \sqrt{{\sqrt{1+4\tau^2} - 1 \over 2 \sqrt{1+4\tau^2}}},
\end{equation}
and
\begin{subequations}
\begin{align}
\phi_{a}(t) =& \dfrac{\phi_\omega(t)-\phi_\omega(0)}{2} - \tan^{-1}\left({2\tau \over \sqrt{2+4\tau^2}}\right) \nonumber \\
&+ {i \over \sqrt{2}} \text{EllipticE}[i \sinh^{-1}(2\tau), 1/2], \\
\phi_{b}(t) =& \dfrac{\phi_\omega(t)+\phi_\omega(0)}{2} - \tan^{-1}\left({2\tau \over \sqrt{2+4\tau^2}}\right) \nonumber \\
&- {i \over \sqrt{2}} \text{EllipticE}[i \sinh^{-1}(2\tau), 1/2] - \dfrac{\pi}{2},
\end{align}
\end{subequations}
with $\text{EllipticE}(\phi,m)=\int_0^\phi [1-m\sin^2(\theta)]^{1/2} d\theta$.
It is interesting to consider a simple case in which $|\omega(t)|=const.=|\omega_0|$.
In this instance we have such a situation that $P_+^-(t)=|b(t)|^2$ $(P_+^+(t)=|a(t)|^2)$ goes from 0 (1), at $t=0$, to ${1 / 2}$ $\bigl( {1 / 2} \bigr)$, when $t \rightarrow \infty$, as it can be appreciated by their plots in Fig. \ref{fig:TPB1}: full blue and dashed red lines, respectively.
In Fig. \ref{fig:DC1} we may appreciate the time behaviour of $\hbar\Delta(t)/|\omega_0|$ related to this specific physical scenario.
This specific out of resonance time-dependent scenario, then, asymptotically evolves the initial state $\ket{+}$ towards an equal-weighted superposition of the two eigenstates of $\hat{\sigma}^z$.
One can convince oneself that this circumstance is intimately related to the fact that, in this case, the ``detuning'' $\Delta(t)$ vanishes asymptotically (see Fig. \ref{fig:DC1}), getting then established a dynamical reply of the system as if it were under the generalized resonance condition.

\subsubsection{CASE 2}\label{Case 2}
The second scenario is based on the following assumption
\begin{align}
\Theta(t) = 2 \tan^{-1}\left({\tau \over \sqrt{2+\tau^2}}\right),
\end{align}
which, notwithstanding its apparent similarity with the previous case given in Eq. \eqref{Theta 1}, leads however to a remarkable different temporal behaviour of the correspondent generalized Rabi system.
This time it results in
\begin{equation} \label{Position 4}
\int_0^t{|\omega(t')|\over\hbar}\cos[\Theta(t')] dt'= \tan^{-1}(\tau),
\end{equation}
so that the solutions of \eqref{LC problem} read
\begin{equation}\label{a b case 4}
\begin{aligned}
|a(t)| = {1 \over \sqrt{1+\tau^2}}, \qquad
|b(t)| = {\tau \over \sqrt{1+\tau^2}} = |a(t)|\tau,
\end{aligned}
\end{equation}
and
\begin{subequations}
\begin{align}
\phi_{a}(t) =& \dfrac{\phi_\omega(t)-\phi_\omega(0)}{2} - \tan^{-1}\left({\tau \over \sqrt{2+\tau^2}}\right) \nonumber \\
&- {1 \over 2} \biggl[ {\tau \sqrt{2+\tau^2} \over 2} + \sinh^{-1}\left({\tau \over \sqrt{2}}\right) \biggr], \\
\phi_{b}(t) =& \dfrac{\phi_\omega(t)+\phi_\omega(0)}{2} - \tan^{-1}\left({\tau \over \sqrt{2+\tau^2}}\right) \nonumber \\
&+ {1 \over 2} \biggl[ {\tau \sqrt{2+\tau^2} \over 2} + \sinh^{-1}\left({\tau \over \sqrt{2}}\right) \biggr] - \dfrac{\pi}{2}.
\end{align}
\end{subequations}
Finally, the special form of $\Delta(t)$ underlying this specific scenario is
\begin{equation} \label{Omega case 4}
\Delta(t) = \left[ {2+(1-\tau^2)(2+\tau^2) \over 2(1+\tau^2)\sqrt{2+\tau^2}}\right] {|\omega(t)| \over \hbar}.%\equiv \tilde{\beta(\tau)}|\omega(t)|=\beta(t)|\omega(t)|
\end{equation}
In this case, it is easy to see that if $|\omega|=const.=|\omega_0|$, $P_+^-(t)=|b(t)|^2$ ($P_+^+(t)=|a(t)|^2$) goes from 0 (1) to 1 (0) asymptotically.
These behaviours, evoking the transition probabilities in the Landau-Zener scenario \cite{LZ}, are illustrated by full blue and dashed red lines, respectively, in Fig. \ref{fig:TPB2}.
%\begin{figure}[htbp]
%\centering
%\includegraphics[scale=0.8]{ModsaAndbCase3and4.eps}
%\caption{(Color online) The dynamical behaviour of $|a(\tau)|$ and $|b(\tau)|$ with $\tau =|\omega_0|t/\hbar$.
%Full blue and dashed green lines correspond to $|a(\tau)|$ for cases 1 and 2, respectively.
%Analogously, dot-dashed red and dotted brown lines correspond to $|b(\tau)|$.}
%\label{fig:FG02}
%\end{figure}
In this case, the time behaviour of the ``detuning'' $\hbar\Delta(t)/|\omega_0|$ is characterized by an asymptotic linear dependence on $t$, as shown in Fig. \ref{fig:DC2}.
As in the resonant scenario, even here different time-dependences of the magnetic field may give rise to qualitatively different time evolutions of $P_+^-(t)$ with respect to the Rabi scenario.

As a final remark we want to emphasize that it could be very hard to get analytical expressions for $a(t)$ and $b(t)$, in Eq. \eqref{Gen a} and \eqref{Gen b}, respectively, depending on the choice of $\Theta(t)$ and two of the three Hamiltonian parameters.
Nevertheless, such a bottleneck does not influence our capability to predict the Rabi transition probability and the expression of $\Delta(t)$ in order to know how to engineer the magnetic fields to get the desired time evolution.
Indeed, we would be always able to find accordingly the expressions of $|a(t)|$ and $|b(t)|$.
Thus, as a consequence, when $\Omega(t)$, $|\omega(t)|$ and $\dot{\phi}_\omega(t)$ are chosen in such a way to generate the same detuning $\Delta(t)$, the related different physical scenarios share the same analytical expressions of $|a(t)|$ and $|b(t)|$ and then of all physical observables depending only on these quantities, e.g. $P_+^-(t)$ or
\begin{equation}
\average{\pm|U^\dagger(t)\hat{\sigma}^zU(t)|\pm}= \pm {\hbar} (|a(t)|^2-|b(t)|^2).
\end{equation}

\section{An exotic application}

In this section we present a possible intriguing application of our results in the classical context of guided wave optics.
Let us consider two electromagnetic modes propagating in the same direction (say the $z$ direction) and characterized by the two complex amplitudes $A$ and $B$ \cite{Yariv}.
These may be defined in such a way that their squared modulus, $|A|^2$ and $|B|^2$, represent the power carried by the two modes.
The amplitudes $A$ and $B$ does not depend on the coordinate $z$ if the medium through which the modes propagate is unperturbed.
However, in a more realistic and experimental scenario several causes may perturb the medium, e.g. an electric field, a sound wave, surface corrugations, etc..
In this case, power is exchanged by the two modes and the amplitudes result mutually coupled in accordance with the following equations \cite{Yariv}
\begin{equation}\label{Problem electric modes}
\begin{aligned}
{dA(z) \over dz}&=k_{ab}(z)e^{-i\Delta z}B(z), \\
{dB(z) \over dz}&=k_{ba}(z)e^{i\Delta z}A(z).
\end{aligned}
\end{equation}
Here $\Delta$ is the phase-mismatch constant and $k_{ab}(z)$ and $k_{ba}(z)$ are complex coupling coefficients determined by the particular physical situation we analyse.
The power conservation may be written as  $\dfrac{d}{dz}(|A(z)|^2+|B(z)|^2)=0$ and it is accomplished by the condition $k_{ab}(z)=-k_{ba}^*(z)=k(z)$ \cite{Yariv}.
It is easy to see that, after straightforward algebraic manipulations, the system may be cast in the following form
\begin{equation}\label{su(1,1) problem z dependent}
i{dV(z) \over dz}=H(z)V(z),
\end{equation}
where $V(z)=[\tilde{A}(z),\tilde{B}(z)]^T$, with
\begin{equation}
\tilde{A}(z)=A(z)e^{i\Delta z/2}, \quad \tilde{B}(z)=B(z)e^{-i\Delta z/2},
\end{equation}
and
\begin{equation}\label{Hamiltonian z dependent}
H(z)=
\begin{pmatrix}
-\Delta/2 & \gamma(z) \\
\gamma^*(z) & \Delta/2
\end{pmatrix},
\end{equation}
having put $\gamma(z) \equiv ik(z)$.

We see immediately that the problem under scrutiny is mathematically equivalent to a Schr\"odinger dynamical problem based on a $su$(2) Hamiltonian.
The physical relevance of our results in connection with this classical guided wave optics scenario may be easily clarified.
Writing $V(z)=\mathcal{U}(z)V(0)$, then the system reads
\begin{equation}\label{U(z) problem}
i{d\mathcal{U}(z)\over dz}=H(z)\mathcal{U}(z), \quad \mathcal{U}(0)=\mathbb{1},
\end{equation}
being nothing but the problem for which we have sets of exact solutions related to specific relations between the Hamiltonian parameters exposed in Sec. \ref{MN Res}.
Therefore, our solvability conditions and the specific cases reported in Sec \ref{Examples}, adapted to the context under scrutiny, furnish special links between $\Delta$ and $k(z)$ turning out in exactly solvable scenarios for the classical problem of two propagating modes in a perturbed medium.

\section{Conclusions}

The Rabi scenario consists in a spin-1/2 subjected to a time-dependent magnetic field precessing around the quantized axis ($\hat{z}$) \cite{Rabi 1937} and is characterized by three time-independent parameters: $\Omega_0$, $|\omega_0|$ and $\dot{\phi}_0$.
Rabi shows that when $\Omega_0+\hbar\dot{\phi}_0/2=\Delta=0$ the transverse magnetic field acts as a probe of the energy separation $2\Omega_0$ due to the longitudinal field alone.
The measurable physical quantity revealing $\Omega_0$ is the transition probability $P_+^-(t)=\average{-|U(t)|+}$ which, at resonance, oscillates between 0 and 1 with frequency now referred to as Rabi frequency.

In this work we generalize this Rabi scenario by assuming an SU(2) general time-dependent Hamiltonian model where then $\Omega_0$, $|\omega_0|$ and $\dot{\phi}_0$ are now replaced with time-dependent counterparts.
Along the lines of the Rabi approach \cite{Rabi 1954}, we firstly show that, in the rotating frame with the time-dependent angular frequency $\dot{\phi}_\omega(t)$, the condition $\Omega(t)+\hbar\dot{\phi}_\omega(t)/2=\Delta(t)=0$ plays the same role of the Rabi resonance condition in the Rabi scenario.
Such an occurrence makes of basic interest a direct comparison between the Rabi scenario and its generalized version on both time-dependent resonance and out of resonance ($\Delta(t) \neq 0$) cases.
To bring to light the occurrence of analogies and differences, we have focussed our attention on the study of the transition probability $P_+^-(t)$ between the two eigenstates of $\hat{S}^z$.

We show that, on resonance, $P_+^-(t)$ depends only on the integral of $|\omega(t)|$.
Our examples illustrate that this circumstance determines a transition probability characterized by three possible different regimes: oscillatory (the only one dominating the Rabi scenario), monotonic and mixed which means an initial oscillatory transient followed by an asymptotic monotonic behaviour.

To capture significant dynamical consequences stemming from the detuning time dependence, we have constructed \cite{Mess-Nak} exactly solvable problems and analysed the corresponding quantum dynamics of the spin-1/2.
We have thus highlighted that when $\Delta(t)$ is proportional to $|\omega(t)|$, the main effect emerging in the time behaviour of $P_+^-(t)$ is a scale effect both in amplitude and in frequency (like in the Rabi scenario).
We have further investigated two specific exactly solvable scenarios of experimental interest for which $\Delta(t)/|\omega(t)|$ varies over time.
One of them predicts a Landau-Zener transition, while the other an equal weighted superposition of the two states of the system.
It is important to underline that our examples illustrate exactly solvable cases for which, then, the corresponding system dynamics may be fully disclosed.
We highlighted, however, that when one is interested in the Rabi transition probability $P_+^-(t)$ only, it is enough the knowledge of $|a(t)|$ and $|b(t)|$.
We point out that this circumstance leads us to wider and richer classes of physical scenarios, since it gets us rid of possible analytical difficulties stemming from Eq. \eqref{Integral R}.

We underline that the knowledge of the exactly solvable problems reported in this paper provide stimulating ideas for technological applications with single qubit devices.
In addition it furnishes ready-to-use tools for interacting qudits systems \cite{GMN, GMIV, GBNM}, being of relevance in several fields, from condensed matter physics \cite{Calvo, Borozdina} to quantum information and quantum computing \cite{Petta, Anderlini, Foletti, Das Sarma Nat}.

As a conclusive remark, we emphasize the applicability of our approach to the propagation of two electromagnetic modes in a perturbed medium.
We have shown the mathematical equivalence of this problem to that of a 2x2 $su(2)$ dynamical problem.
As a consequence, the main results achieved in our paper could be advantageously adapted to a guided wave optics scenario.

\section*{Acknowledgements}

ASM de Castro acknowledges the Brazilian agency CNPq financial support Grant No. 453835/2014-7.
RG acknowledges for economical support by research funds difc 3100050001d08+ (University of Palermo) in memory of Francesca Palumbo.


\begin{thebibliography}{00}

\bibitem{Hioe}
Hioe  F T 1987 J. Opt. Soc. Am. B \textbf{4} 1327 .

\bibitem{Haroche}
Haroche S and Raimond J M 2006 \textit{Exploring the Quantum: Atoms Cavities And Photons}, (Oxford: Oxford University Press).

\bibitem{Daems}
Daems D, Ruschhaupt A, Sugny D and Guerin S 2013 Phys. Rev. Lett. \textbf{111} 050404.

\bibitem{Greilich}
Greilich A, Economou S E, Spatzek S, Yakovlev D R, Reuter D, Wieck A D, Reinecke T L and Bayer M 2009 Nature Phys. \textbf{5} 262.

\bibitem{NC}
Nielsen M A and Chuang I L 1990  \textit{Quantum Computation and Quantum Information},  (Cambridge: Cambridge University Press).

\bibitem{Braak}
 Braak D,  Chen Q -H, Batchelor M T and Solano E 2016 J. Phys. A: Math. Theor. \textbf{49} 300301.

\bibitem{Oliveira}
 Oliveira I S, Bonagamba T J, Sarthour R S, Freitas J C C and deAzevedo E R  2007 NMR Quantum Information Processing, Elsevier.

\bibitem{Rabi 1937}
Rabi I I 1937 Phys. Rev. \textbf{51} 652.

\bibitem{Rabi 1954}
Rabi I I, Ramsey N F and Schwinger J 1954 Rev. Mod. Phys. \textbf{26} 167.

\bibitem{Schwinger 1937}
Schwinger J 1937 Phys. Rev. \textbf{51} 648.

\bibitem{Vandersypen}
Vandersypen L M K and Chuang I L 2004 Rev. of Mod. Phys. \textbf{76} 1037.

\bibitem{JC}
Jaynes E T and Cummings F W 1963 Proc. IEEE \textbf{51} 89.

\bibitem{Issue}
Braak D \textit{et al}. 2016 J. Phys. A: Math. Theor. \textbf{49} 300.

%\bibitem{Sakurai}
%J. J. Sakurai, \textit{Modern Quantum Mechanics}, 1994 (Reading, MA: Addison-Wesley).
%
%\bibitem{Bloch-Siegert 1940}
%F. Bloch and A. J. Siegert, Phys. Rev. \textbf{57}, 522 (1940).
%
%\bibitem{Bloch-Alvarez 1940}
%L. Alvarez and F. Bloch, Phys. Rev. \textbf{57}, 111 (1940).
%
%\bibitem{Bloch 1946}
%F. Bloch, Phys. Rev. \textbf{70}, 460 (1946).
%
%\bibitem{Bloch-Wangsness 1953}
%R. K. Wangsness and F. Bloch, Phys. Rev. \textbf{89}, 728 (1953).

\bibitem{Bagrov}
Bagrov V G , Gitman D M , Baldiotti M C and Levin A D 2005 Annalen der Physik \textbf{14} (11) 764.

\bibitem{KunaNaudts}
Maciej K and Naudts J 2010 Reports on Mathematical Physics \textbf{65} (1) 77.

\bibitem{Das Sarma}
Barnes E and Das Sarma S 2012 Phys. Rev. Lett. \textbf{109} 060401.

\bibitem{Mess-Nak}
Messina A and Nakazato H  2014 J. Phys. A: Math. Theor. \textbf{47} 445302.

\bibitem{MGMN}
Markovich L A, Grimaudo R, Messina A and Nakazato H 2017 Annals of Physics \textbf{385} 522.

\bibitem{GMN}
Grimaudo R, Messina A and Nakazato H  2016 Phys. Rev. A \textbf{94} 022108.

\bibitem{GMIV}
Grimaudo R, Messina A, Ivanov P  A and Vitanov N V 2017 J. Phys. A \textbf{50} 175301.

\bibitem{GBNM}
Grimaudo R, Belousov Y, Nakazato H and Messina A, Annals of Physics \textbf{392}, 242 (2017).

\bibitem{Calvo}
Calvo R, Abud J E, Sartoris R P and Santana R C 2011 Phys. Rev. B \textbf{84} 104433 .

\bibitem{Borozdina}
Borozdina Y B,  Mostovich E,  Enkelmann V,  Wolf B,  Cong P T,  Tutsch U,  Lang M and Baumgarten M 2014 J. Mater. Chem. C \textbf{2} 6618.

\bibitem{Petta}
Petta J R  \textit{et al} 2005 Science \textbf{309} (5744) 2180.

\bibitem{Anderlini}
Anderlini M, Lee P J, Brown B L, Sebby-Strabley J, Phillips W D, and Porto J V 2007 Nature \textbf{448} (7152) 452.

\bibitem{Das Sarma Nat}
Wang Xin , Bishop L S, Kestner J P, Barnes E, Sun Kai and  Das Sarma S 2012 Nat. Comm. \textbf{3} 997.

\bibitem{Foletti}
Bluhm H, Foletti S, Neder I, Rudner M, Mahalu D, Umansky V and Yacoby A 2011 Nat. Phys. \textbf{7} 109.

\bibitem{LZ}
Landau L  1932 Phys. Z. Sowjetunion \textbf{2} 46; Zener C 1932 Proc. R. Soc. A \textbf{137} 696.

\bibitem{Yariv}
A. Yariv, IEEE J. Quantum Electron. \textbf{9}, 9 (1973).

\end{thebibliography}
\end{document}